\documentclass[aps,pra,twocolumn,showpacs,superscriptaddress,amssymb,amsmath,nobibnotes]{revtex4-1}
\usepackage{graphicx}
\usepackage{hyperref}
\usepackage{comment}
\usepackage{multirow}
\usepackage{float}
\usepackage{physics}
\usepackage{bm}
\usepackage{times}

\hypersetup{%
   pdfpagemode=None, 
   pdfstartpage=1,
   pdfmenubar=true,
   pdftoolbar=true,
   colorlinks = true,
   linkcolor=blue,
   citecolor=blue,
   urlcolor=blue,
   bookmarksopen=false
}

\begin{document}

\title{Van der Waals coefficients for interactions of dysprosium and erbium atoms with alkali-metal and alkaline-earth-metal atoms}

\author{Klaudia Zaremba-Kopczyk}
\affiliation{Faculty of Physics, University of Warsaw, Pasteura 5, 02-093 Warsaw, Poland}
\author{Micha{\l} Tomza}
\affiliation{Faculty of Physics, University of Warsaw, Pasteura 5, 02-093 Warsaw, Poland}
\author{Maxence Lepers}
\email{maxence.lepers@u-bourgogne.fr}
\affiliation{Laboratoire Interdisciplinaire Carnot de Bourgogne, CNRS, Universit\'{e} de Bourgogne Franche-Comt\'{e}, 21078 Dijon, France}
\date{\today}

\begin{abstract}

The long-range part of the interatomic interactions plays a substantial role in the collisional dynamics of ultracold gases. Here, we report on the calculation of the isotropic and anisotropic $C_6$ coefficients characterizing the van der Waals interaction between dysprosium or erbium atoms in the two lowest energy levels and the ground-state alkali-metal (Li, Na, K, Rb, Cs, Fr) or alkaline-earth-metal (Be, Mg, Ca, Sr, Ba) atoms. The calculations are done using the integral of dynamic dipole polarizabilities at so-called imaginary frequencies of the two interacting atoms. For all atom pairs, we find that the isotropic $C_6$ coefficients are two or three orders of magnitude larger than the anisotropic ones. Those coefficients are essential for modeling collisional properties of heteronuclear quantum mixtures containing highly magnetic dysprosium or erbium atoms and alkali-metal or alkaline-earth-metal atoms.

\end{abstract}

\maketitle

\section{Introduction}

Dipolar quantum gases have been experiencing a surge in interest over the last years, driven by the experimental breakthroughs in reaching quantum degeneracy with ultracold gases of highly magnetic atoms~\cite{GriesmaierPRL05,LuPRL11,LuPRL12,AikawaPRL12,AikawaPRL14,NaylorPRA15,MiyazawaPRL22} and continuous advances in the production of ultracold polar molecules~\cite{GadwayJPhysB16,MosesNatPhys17}. Ultracold gases composed of particles possessing a large intrinsic magnetic or/and electric dipole moment are characterized by the unique combination of tunable short-range contact interactions and long-range anisotropic dipole-dipole interactions, offering exceptional controllability with external electromagnetic fields. This feature of dipolar quantum gases has opened up new possibilities for exploring few-body and many-body physics of strongly correlated systems~\cite{BaranovCR12,ChomazRPP23}, controlled chemistry~\cite{deMirandaNatPhys11,BohnScience17}, quantum information~\cite{DeMillePRL02,NiCS18}, and physics beyond the Standard Model~\cite{SafronovaRMP18}. 

Numerous fascinating phenomena have already been observed with dipolar gases composed of highly magnetic lanthanide dysprosium and erbium atoms, just to mention the chaotic spectra of Feshbach resonances~\cite{FrischNature14}, Fermi surface deformation~\cite{AikawaScience14}, and quantum-stabilized states -- self-bound droplets~\cite{KadauNature16,SchmittNature16} and supersolids~\cite{TanziPRL19,BoettcherPRX19,ChomazPRX19}. While the experimental studies of ultracold polar molecules have been focused so far mainly on heteronuclear bialkali molecules~\cite{NiScience08,OspelkausScience10,MolonyPRL14,ParkPRL15,GuoPRL16,YeSciAdv18}, the production of dimers possessing much more complex internal structure, such as Er$_2$~\cite{FrischPRL15} and DyK~\cite{Soave2023}, has also been demonstrated. The ongoing advances in the production and manipulation of ultracold dipolar molecules hold promise for the realization of novel exotic states of quantum matter, like molecular superfluids and supersolids~\cite{GorshkovPRL11,FedorovSciRep16,SchmidtPRR22}.

Recently, there has been a growing interest in degenerate mixtures containing highly magnetic atoms, such as Cr ($^7S_3$; magnetic dipole moment of 6 Bohr magnetons, $\,\mu_B$), Eu ($^8S_{7/2}$; 7$\,\mu_B$), Er ($^3H_6$; 7$\,\mu_B$), Ho ($^4I^{o}_{15/2}$; 9$\,\mu_B$), or Dy ($^5I_8$; 10$\,\mu_B$), as they offer great versatility in exploring novel physical phenomena. Heteronuclear molecules formed via magneto- or photoassociation will possess large both electric and magnetic dipole moments, combining strong anisotropic interactions of both electric and magnetic nature with the complexity of molecular electronic structure. Aside from the formation of molecules in non-trivial electronic states, such heteronuclear mixtures can be employed in studies of polaron physics in systems with dominant dipolar interactions~\cite{KainPRA14,WenzelPS18,ArdilaJPhysB19}, Efimov physics~\cite{PiresPRL14}, exotic Fulde-Ferrell-Larkin-Ovchinnikov states in systems with significant mass imbalance~\cite{RadzihovskyRPP10,WangSciRep17}, and binary supersolids~\cite{BlandPRA22}.  With current experiments on degenerate mixtures of Dy and K atoms~\cite{RavensbergenPRA18,RavensbergenPRL20,YePRA22}, Dy and Er atoms~\cite{TrautmannPRL18,DurastantePRA20,PolitiPRA22}, Er and Li atoms~\cite{SchaeferPRA22}, Cr and Li atoms~\cite{CiameiPRA22,CiameiPRL22}, and Er and Yb atoms~\cite{SchaeferPRA23}, the realization of theoretical proposals is becoming more and more feasible. Therefore, the electronic structure of molecules containing highly magnetic transition-metal and lanthanide atoms, such as Cr--~\cite{PavlovicPRA10, TomzaPRA13} and Eu--alkali-metal and alkaline-earth-metal dimers~\cite{TomzaPRA14}, have been theoretically investigated alongside the collisional properties of ultracold heteronuclear mixtures: Cr+Li~\cite{Zaremba23}, Cr+Rb~\cite{PavlovicPRA10}, Cr+Ca$^+$/Sr$^+$/Ba$^+$/Yb$^+$~\cite{TomzaPRA15}, Eu+Li/Rb~\cite{ZarembaPRA18}, Er+Li~\cite{GonzalezMartinezPRA15}, Er+Yb~\cite{KosickiNJP20,FryePRX20}, Er+Sr~\cite{FryePRX20}, and Dy+Sr/Yb~\cite{FryePRX20}. Additionally, the \textit{ab initio} studies of interatomic interactions in homonuclear dimers of highly magnetic lanthanide atoms such as Eu$_2$~\cite{BuchachenkoJCP09}, Er$_2$~\cite{TiesingaNJP21}, and Tm$_2$~\cite{TiesingaNJP21} have also been reported.

Despite the significant increase in computational power and development of computational methods for electronic structure calculations over the last decades, a full \textit{ab initio} approach to characterize the interatomic interactions in dimers containing heavy atoms in non-trivial electronic states would require the use of an enormous active space to account for all possible electron configurations, and that far exceeds currently available computational resources. In the case of lanthanide atoms, the unpaired electrons occupying the $4f$ or $5d$ shells, submerged under a closed $6s$ shell, give rise to large magnetic moments and large electronic orbital angular momenta of the atoms, which in turn lead to anisotropic interatomic interactions. Large basis sets would need to be employed in electronic structure calculations for dimers involving these atoms to ensure proper description of interactions at large internuclear distances. It is, therefore, impossible to compute Born-Oppenheimer potential-energy curves covering the full range of internuclear distances with the accuracy needed to precisely predict the scattering properties in ultracold systems containing highly magnetic lanthanides.

At ultralow temperatures, neutral atoms interact mainly via short-range van der Waals (vdW) interactions, whose leading term scales as $1/R^6$ with internuclear distance $R$, and, for atoms with magnetic moments, also via long-range and anisotropic magnetic dipole-dipole interactions that scale as $1/R^3$. In the case of atoms whose electronic state is not spherically symmetric, the vdW interactions are also anisotropic. This anisotropy of interactions induces couplings between the scattering states in the open channels and bound molecular states in the closed channels, significantly modifying the collisional properties of an ultracold quantum gas containing magnetic atoms in non-$S$ states. Due to the large number of scattering channels involved, a complete coupled-channels approach to quantum scattering calculations would be extremely computationally demanding. Therefore, simplified models for the ultracold atom-atom collisions have been developed~\cite{ChinRMP10}. Since the tail of the interaction potential plays the most significant role in the two-body dynamics, it is crucial to know accurate values of the van der Waals (dispersion) $C_6$ coefficients that enter the leading term of the multipole expansion, $-C_6/R^6$, while the effect of short-range spin-exchange interactions can be included within the phase of the scattering wave function.

The aim of the present study is to compute the leading van der Waals coefficients, $C_6$, for Dy and Er atoms in their two lowest electronic states interacting with ground-state alkali-metal (Li, Na, K, Rb, Cs, Fr) and alkaline-earth-metal (Be, Mg, Ca, Sr, Ba) atoms, and to establish a general computational scheme applicable to other similar systems. To this end, we employ the sum-over-states method to calculate the dynamic electric dipole polarizabilities, which are further used to compute the $C_6$ coefficients with the Gaussian quadrature method. We derive the formulas for the $C_6$ coefficients in the basis of both fine and hyperfine atomic levels and present the numeric values of the isotropic $C_{6,00}$ and anisotropic $C_{6,20}$ coefficients.

The outline of this paper is as follows. In Section~\ref{sec:theory}, we introduce the electronic structure of the considered atoms, define the dynamic electric dipole polarizability, recall the formula for the second-order energy correction resulting from the vdW interactions between two neutral atoms in the presence of both fine and hyperfine interactions and, finally, we provide the formulas for the $C_6$ coefficients: isotropic $C_{6,00}$ and anisotropic $C_{6,20}$. In Section~\ref{sec:results}, we present the computed values of $C_{6,00}$ and $C_{6,20}$ coefficients for Dy/Er+alkali-metal/ alkaline-earth-metal atom pairs and discuss the obtained results. Section~\ref{sec:summary} contains a summary of our findings and concluding remarks.

\section{Methodology}
\label{sec:theory}

\subsection{Electronic structure}

The electronic configuration of ground-state dysprosium Dy($^5I_8$) is [Xe]$4f^{10}6s^2$ with total electronic angular momentum $J=8$, orbital angular momentum $L=6$ and spin angular momentum $S=2$. The first excited level Dy($^5I_7$) with $J=7$ has the same electronic configuration and belongs to the same LS manifold as the ground state. The lowest electronic configuration of erbium is [Xe]$4f^{12}6s^2$ with the ground state Er($^3H_6$) ($J=6$, $L=5$, $S=1$) and first excited state Er($^3F_4$) ($J=4$, $L=3$, $S=1$). Ground-state alkali-metal atoms (AMs) and ground-state alkaline-earth-metal atoms (AEMs) possess a much simpler electronic structure, with spherically-symmetric ground states described by $^2S_{1/2}$ ($J=1/2$, $L=0$, $S=1/2$) and $^1S_{0}$ ($J=0$, $L=0$, $S=0$) terms, respectively. The total angular momenta $J$ and their projections on the quantization axis $M$ are good quantum numbers in the presence of a spin-orbit coupling, and we use them to label the atomic energy levels throughout our derivations in Sec.~\ref{sec:theory}~B~--~D.

\subsection{Long-range potential energy}

We consider two distant charge distributions $A$ and $B$, whose centers of mass are separated by distance $R$, in a referential frame whose $z$ axis points from $A$ to $B$. In spherical coordinates and in atomic units (used throughout this paper), the multipolar expansion of their interaction energy can be written as \cite{stone1996, kaplan2006, groenenboom2007, lepers2018}:
\begin{equation}
\label{eq:VAB}
  V_\mathrm{AB}(R) = \sum_{l_A,l_B=0}^{+\infty}  \sum_{m=-l_<}^{+l_<} \frac{f_{l_A l_B m}}{R^{1+l_A+l_B}} Q_{l_A, m}(A) Q_{l_B, -m}(B) \, ,
\end{equation}
where $l_A$ and $l_B$ describe the tensor rank related to the multipole moments $Q_{l_A, m}(A)$ and $Q_{l_B, -m}(B)$ of the charge distributions $A$ and $B$, respectively, and $-l_< \leq m \leq l_<$, where $l_< = \min(l_A, l_B)$. The number factor $f_{l_A l_B m}$ equals to:
\begin{equation}
  f_{l_A l_B m} =(-1)^{l_B} \sqrt{\frac{(2 l_A +2 l_B)!}{(2 l_A)! (2 l_B)!}} C^{l_A+l_B,0}_{l_A m l_B -m} \, 
\end{equation}
with $C^{a \alpha}_{b \beta c \gamma}$ denoting a Clebsch-Gordan (CG) coefficient.

If we assume that $A$ and $B$ are two atoms, and $B$ is in an $S$ state, the first-order correction of the multipolar expansion of Eq.~\eqref{eq:VAB} is equal to zero. The matrix element describing the second-order energy correction reads \cite{LiJPhysB16}:
\begin{equation}
\label{Vmatrix}
  \begin{aligned}
  & \langle \beta_A J_A M_A' \beta_B J_B M_B' | \hat{V}^{(2)}_\mathrm{AB} | \beta_A J_A M_A \beta_B J_B M_B \rangle
  \\ & =-\sum_{l_A l_B l_A' l_B' } \frac{(-1)^{l_B + l_B' + 2 J_A + 2 J_B}}{R^{2+l_A+l_B+l_A'+l_B'}} 
  \\ & \cross\sqrt{\frac{(2 l_A + 2 l_B +1)! (2 l_A' + 2 l_B' +1)!}{(2 l_A)! (2 l_B)!(2 l_A')! (2 l_B')!}} 
  \\ & \cross\sum_{k_A k_B k q} (-1)^{k_A +k_B} (2 k_A +1) (2 k_B +1) 
  \\ & \cross C^{k 0}_{l_A+l_B,0,l_A'+l_B',0} C^{k 0}_{k_A q k_B -q} 
  \\ & \cross \begin{Bmatrix}
    l_B' & l_B & k_B \\
    l_A' & l_A & k_A \\
    l_A'+l_B' & l_A+l_B  & k
  \end{Bmatrix} \sum_{\beta_A'' J_A'' \beta_B'' J_B''} \\ &\cross \frac{\langle\beta_A J_A ||\hat{Q}_{l_A }||\beta_A'' J_A'' \rangle \langle\beta_A'' J_A'' ||\hat{Q}_{l_A'}||\beta_A J_A \rangle }{E_{\beta_A'' J_A''} - E_{\beta_A J_A}+E_{\beta_B'' J_B''} - E_{\beta_B J_B}} 
  \\ & \cross \langle\beta_B J_B||\hat{Q}_{l_B}||\beta_B'' J_B''\rangle \langle\beta_B'' J_B'' ||\hat{Q}_{l_B' }||\beta_B J_B\rangle 
  \\ & \cross
  \begin{Bmatrix}
    l_A' & l_A & k_A \\
    J_A & J_A & J_A'' 
  \end{Bmatrix} \begin{Bmatrix}
    l_B' & l_B & k_B \\
    J_B & J_B & J_B'' 
  \end{Bmatrix} \\&\cross \frac{C^{J_A M_A'}_{J_A M_A k_A q} C^{J_B M_B'}_{J_B M_B k_B, -q}}{\sqrt{(2 J_A+1) (2 J_B+1)}} \, ,
  \end{aligned}
\end{equation}
where the subscripts denote values corresponding to atoms $A$ and $B$, respectively, $E_{\beta_{\{A,B\}} J_{\{A,B\}}}$ ($E_{\beta_{\{A,B\}}'' J_{\{A,B\}}''}$) is the energy of level $|\beta_{\{A,B\}} J_{\{A,B\}}\rangle$ ($|\beta_{\{A,B\}}'' J_{\{A,B\}}''\rangle$) ($\beta$ denotes all remaining quantum numbers describing the state of an atom), and $\langle\beta_{\{A,B\}} J_{\{A,B\}}||\hat{Q}_{l_{\{A,B\}}}||\beta_{\{A,B\}}'' J_{\{A,B\}}''\rangle$ is the reduced transition multipole moment between $|\beta_{\{A,B\}} J_{\{A,B\}}\rangle$ and $|\beta_{\{A,B\}}'' J_{\{A,B\}}''\rangle$ levels. The selection rules impose that $M_A + M_B = M_A' + M_B'$. The pairs ($k_A$, $k_B$) and the value of $k$ are constrained by the values of ($l_A$, $l_A'$, $l_B$, $l_B'$) and define the possible ranks of the tensorial terms; $q$ is limited by the minimum of $k_A$ and $k_B$~\cite{LepersPCCP11}. The first curly brackets contain a Wigner 9-j symbol, whereas the latter two contain a Wigner 6-j symbol.

In this paper, we consider the induced-dipole--induced-dipole interaction term ($l_A = l_B = l_A' = l_B' = 1$) and neglect higher-order terms as they decay faster than $R^{-6}$. In our particular case, atom $A = \{ \text{Dy}(^5I_8),\, \text{Dy}(^5I_7),\, \text{Er}(^3H_6),\, \text{Er}(^3F_4)\}$ and atom $B = \{\text{AM}(^2S_{1/2}),\, \text{AEM}(^1S_0)\}$. Therefore, $k_A = \{0,2\}$ and $k_B=0$, which implies that $k = k_A = \{0,2\}$ and $q = 0$. The CG coefficient $C^{\alpha a }_{\alpha a 0 0}$ equals to~1. With the above assumptions, the matrix element from Eq.~(\ref{Vmatrix}) reads:
\begin{equation}
\label{VDyErX}
  \begin{aligned}
  & \langle \beta_A J_A M_A' \beta_B J_B M_B' | \hat{V}_\mathrm{AB}^{(2)} | \beta_A J_A M_A \beta_B J_B M_B \rangle \\
  & = -\frac{30}{R^6} \sum_{\beta_A'' J_A'' \beta_B'' J_B''} \frac{(-1)^{J_A + J_A''} (-1)^{J_B + J_B''}}{\sqrt{(2 J_A+1) (2 J_B+1)}} \\ 
  & \cross \frac{ |\langle\beta_A J_A ||\hat{Q}_1||\beta_A'' J_A'' \rangle|^2   |\langle\beta_B J_B ||\hat{Q}_1||\beta_B'' J_B''\rangle| }{E_{\beta_A'' J_A''} - E_{\beta_A J_A}+E_{\beta_B'' J_B''} - E_{\beta_B J_B}} \\ &\cross \sum_{k_A = 0, 2} (2 k_A +1) C^{k_A 0}_{2 0 2 0} C^{J_A M_A'}_{J_A M_A k_A 0}
  \\ & \cross \begin{Bmatrix}
    1 & 1 & 0 \\
    1 & 1 & k_A \\
    2 & 2 & k_A
  \end{Bmatrix} 
  \begin{Bmatrix}
    1   & 1   & k_A \\
    J_A & J_A & J_A'' 
  \end{Bmatrix} \begin{Bmatrix}
    1   & 1   & 0\\
    J_B & J_B & J_B'' 
  \end{Bmatrix} \\ &\cross  \delta_{M_A, M_A'} \delta_{M_B, M_B'} \, ,
  \end{aligned}
\end{equation}
which can be written as $-C_6(M_A)/R^6$ with the leading $M_A$-dependent van der Waals $C_6$ coefficient equal to:
\begin{equation}\label{c6}
  \begin{aligned}
  & C_6(M_A) = 30 \sum_{\beta_A'' J_A'' \beta_B'' J_B''} \frac{(-1)^{J_A + J_A''} (-1)^{J_B + J_B''}}{\sqrt{(2 J_A+1) (2 J_B+1)}} \\ 
  & \cross \frac{ |\langle\beta_A J_A ||\hat{Q}_1||\beta_A'' J_A'' \rangle|^2   |\langle\beta_B J_B ||\hat{Q}_1||\beta_B'' J_B''\rangle|^2 }{E_{\beta_A'' J_A''} - E_{\beta_A J_A}+E_{\beta_B'' J_B''} - E_{\beta_B J_B}} \\ &\cross \sum_{k_A = 0, 2} (2 k_A +1) C^{k_A 0}_{2 0 2 0} C^{J_A M_A}_{J_A M_A k_A 0} \\ &\cross \begin{Bmatrix}
    1 & 1 & 0   \\
    1 & 1 & k_A \\
    2 & 2 & k_A
  \end{Bmatrix} 
  \begin{Bmatrix}
    1   & 1   & k_A \\
    J_A & J_A & J_A'' 
  \end{Bmatrix} \begin{Bmatrix}
    1   & 1   & 0 \\
    J_B & J_B & J_B'' 
  \end{Bmatrix} \, .
  \end{aligned}
\end{equation}
Note that in Eq.~\eqref{VDyErX}, we have used the relation $\langle\beta J \|\hat{Q}_1\| \beta'' J'' \rangle = (-1)^{J''-J} \langle\beta'' J'' \|\hat{Q}_1\| \beta J \rangle$ for $A$ and $B$. In our upcoming developments, we will also use \cite{VMK}
\begin{equation}
  \begin{Bmatrix}
    1 & 1 & 0 \\
    J & J & J'' 
  \end{Bmatrix} = \frac{(-1)^{1+J+J''}}{\sqrt{3(2J+1)}} \,.
\end{equation}

In the Dy- and Er-AEM systems, each $C_6(M_A)$ is associated with a Hund's case (c) potential-energy curve (PEC) $\Omega^\sigma$, where $\Omega = M_A$ goes from $-J_A$ to $+J_A$, and $\sigma = +$ for $\Omega = 0$ \cite{chang1967}. In the Dy- and Er-AM systems, each $C_6(M_A)$ is associated with two PECs with $\Omega = M_A \pm 1/2$, going from $-J_A-1/2$ to $J_A+1/2$.

\subsection{Dynamic dipole polarizabilities}

The dynamic electric dipole polarizability describes the dynamical response of an atom to an external oscillating electric field and, when calculated as a function of imaginary frequencies, it can be employed in the calculations of $C_6$ coefficients, as discussed in the next subsection. For an atom in a level $\ket{\beta J}$, the $zz$ component of the dynamic electric dipole polarizability $\alpha_{zz}$ at imaginary frequency $i\omega$ can be written as:
\begin{equation}\label{alphadip}
  \begin{aligned}
  & \alpha_{zz}(i\omega;\beta, J, M) = 2 \sum_{(\beta'' J'') \neq (\beta J)}
  \frac{E_{\beta''J''}-E_{\beta J}}{(E_{\beta''J''}-E_{\beta J})^2+\omega^2} 
  \\ & \cross |\langle\beta J||\hat{Q}_{1}||\beta'' J''\rangle|^2 (-1)^{J+J''} \sum_{k=0,2} 
  \sqrt{\frac{2k+1}{2J+1}}  C^{k 0}_{1 0 1 0} C^{J M}_{J M k 0} \\ &\cross
  \begin{Bmatrix}
    1 & 1 & k \\
    J & J & J'' 
  \end{Bmatrix} = \sum_{k=0,2} \frac{C^{k 0}_{1 0 1 0} C^{J M}_{J M k 0}}{\sqrt{2 J + 1}} \alpha_k(i \omega; \beta, J)\, .
  \end{aligned}
\end{equation}
We can further decompose $\alpha_{zz}$ into isotropic (or scalar), $M$-independent $\alpha^\mathrm{scal}_{\beta, J}$ ($k=0$) and anisotropic $\alpha^\mathrm{aniso}_{\beta, J, M}$ ($k=2$) components, expressed in terms of the coupled polarizabilities $\alpha_k$ as \cite{angel1968, li2017a}:
\begin{equation}
  \label{alphascal}
  \alpha^\mathrm{scal}_{\beta, J}(i \omega) = -\frac{1}{\sqrt{3(2J+1)}} \alpha_0(i \omega; \beta, J) \, ,
\end{equation}
\begin{equation}
  \label{alphaaniso}
  \alpha^\mathrm{aniso}_{\beta, J, M}(i \omega) =  \frac{\sqrt{2}(3 M^2 - J(J+1))}{\sqrt{3 J (J+1)  (2 J +3)(4J^2-1)}} \alpha_2(i \omega; \beta, J) \, ,
\end{equation}
where
\begin{equation}\label{alphak}
  \begin{aligned}
  & \alpha_k(i\omega; \beta, J) = 2 \sqrt{2k+1}  \sum_{(\beta'' J'') \neq (\beta J)}
  \frac{E_{\beta''J''}-E_{\beta J}}{(E_{\beta''J''}-E_{\beta J})^2+\omega^2} 
  \\ & \cross |\langle\beta J||\hat{Q}_{1}||\beta'' J''\rangle|^2  (-1)^{J+J''} 
  \begin{Bmatrix}
    1 & 1 & k \\
    J & J & J'' 
  \end{Bmatrix} \, ;
  \end{aligned}
\end{equation}
$\alpha^\mathrm{aniso}_{\beta, J, M}$ can be further related to the so-called tensor polarizability $\alpha^\mathrm{tens}_{\beta, J}$ in the following way:
\begin{equation}
  \label{alphaaniso2}
  \alpha^\mathrm{aniso}_{\beta, J, M}(i \omega) =  \frac{3 M^2 - J(J+1)}{J(2J-1)} \alpha^\mathrm{tens}_{\beta, J}(i\omega) \, ,
\end{equation}
where
\begin{equation}
  \label{alphatens}
  \alpha^\mathrm{tens}_{\beta, J}(i\omega) = \sqrt{\frac{2 J(2J-1)}{3(J+1)(2J+1)(2J+3)}} \alpha_2(i \omega; \beta, J).
\end{equation}
The anisotropic part of the polarizability is zero when $J < 1$, thus for AMs and AEMs.

As one can tell from Eq.~(\ref{alphadip}), the calculation of the dynamic electric dipole polarizabilities requires an accurate knowledge of transition energies and transition dipole moments. For Dy and Er, those atomic data are computed with the semi-empirical method described in Refs.~\cite{LepersPRA14, LiJPhysB16, BecherPRA18, lepers2023}, which gives a good agreement with experimental polarizabilities at real frequencies (see Sec.~\ref{sec:results}). The polarizability data for the alkali-metal AM$(^2S_{1/2})$ and alkaline-earth-metal AEM$(^1S_0)$ atoms used in this paper were provided by Derevianko \textit{et al.}~in Ref.~\cite{DereviankoTables10}.

\subsection{van der Waals $C_6$ coefficients}

In a similar way to polarizabilities, the $C_6$ coefficients of Eq.~(\ref{c6}) can be written as a sum of an isotropic $C_{6,{00}}$ and anisotropic $\propto C_{6,20}$ contributions, $C_{6,20}$ being the only anisotropic contribution since atom $B$ is spherically symmetric \cite{chu2005, chu2007}, namely:
\begin{equation}\label{C6}
  C_6(M_A) = C_{6,00} + \frac{3 M_A^2 - J_A(J_A+1)}{2 J_A (2 J_A -1)} C_{6,20} \, ,
\end{equation}
where $C_{6,00}$ and $C_{6,20}$ can be conveniently expressed in terms of scalar $\alpha^\mathrm{scal}_{\beta, J}$ and tensor $\alpha^\mathrm{tens}_{\beta, J}$ dynamic polarizabilities at imaginary frequencies. To this end, we apply the residue theorem to Eq.~(\ref{c6}):
\begin{equation}
  \label{res}
  \frac{1}{a+b} = \frac{2}{\pi} \int_{0}^{\infty} \frac{a b}{(a^2+u^2)(b^2+u^2)} \mathrm{d}u
\end{equation}
where $a,b>0$: in our case $a=E_{\beta_A'' J_A''} - E_{\beta_A J_A}$, $b=E_{\beta_B'' J_B''} - E_{\beta_B J_B}$, and $u=\omega$. We find that the isotropic coefficient $C_{6,00}$ can be computed using the integral:
\begin{equation}
  \label{C60}
  C_{6,00}= \frac{3}{\pi} \int_{0}^{\infty} \mathrm{d}\omega \alpha^\mathrm{scal}_{\beta_A,J_A}(i\omega)  \alpha^\mathrm{scal}_{\beta_B,J_B}(i\omega)\, ,
\end{equation}
while the $C_{6,20}$ coefficient is given by:
\begin{equation}
  \label{C62}
  C_{6,20} = \frac{3}{\pi} \int_{0}^{\infty} \mathrm{d}\omega \alpha^\mathrm{tens}_{\beta_A,J_A}(i\omega) \alpha^\mathrm{scal}_{\beta_B,J_B}(i\omega)\,.
\end{equation}
Note that our $C_{6,20}$ coefficient does not have the definition as in Ref.~\cite{chu2005, chu2007}.
Following Ref.~\cite{DereviankoTables10}, we compute the $C_{6,k_A 0}$ coefficients using the 50-point Gauss-Legendre quadrature method:
\begin{equation}\label{C6gauss}
  C_{6,\{0,2\}0} = \frac{3}{\pi} \sum_{\kappa=0}^{50} w_\kappa \alpha^\mathrm{\{scal, tens\}}_{\beta_A,J_A}(i\omega_\kappa) \alpha^\mathrm{scal}_{\beta_B,J_B}(i\omega_\kappa) \, .
\end{equation}
The values of Gaussian quadrature abscissas $\omega_\kappa$ and weights $w_\kappa$ are provided in Ref.~\cite{DereviankoTables10}.

\subsection{Hyperfine structure}

Both dysprosium and erbium possess stable fermionic isotopes: $^{161}$Dy and $^{163}$Dy with nuclear spins $I=5/2$, and $^{167}$Er with nuclear spin $I=7/2$. Due to the coupling between the total electronic angular momentum $\bm{J}$ and the nuclear spin angular momentum $\bm{I}$, the total angular momentum $\bm{F} = \bm{J} + \bm{I}$ needs to be introduced. The associated projections of the angular momenta onto the quantization axis will be denoted as $M$, $M_I$, and $M_F$ for $J$, $I$, and $F$, respectively. The hyperfine levels are labeled in the coupled basis $\ket{\beta J I F M_F} = \sum_{M, M_I} C^{F M_F}_{J M I M_I} \ket{\beta J M} \ket{I M_I}$. 

Assuming that the hyperfine-structure energies are negligible compared to $E_{\beta_A'' J_A''} - E_{\beta_A J_A}$ and $E_{\beta_B'' J_B''} - E_{\beta_B J_B}$, the matrix elements of the second-order operator $V_\text{AB}^{(2)}$ in the coupled basis $\ket{\beta_A J_A I_A F_A M_{F,A} \beta_B J_B I_B F_B M_{F,B}}$ can be expressed as those in the uncoupled one, given in Eq.~\eqref{Vmatrix},
\begin{widetext}
\begin{equation}
  \begin{aligned}
    & \langle \beta_A J_A I_A F_A' M_{F,A}' \beta_B J_B I_B F_B' M_{F,B}' 
    | \hat{V}^{(2)}_\mathrm{AB} 
    | \beta_A J_A I_A F_A M_{F,A} \beta_B J_B I_B F_B M_{F,B} \rangle 
  \\ & = \sum_{M'_A M'_{I,A}} C^{F'_A M'_{F,A}}_{J_A M'_A I_A M'_{I,A}}
         \sum_{M'_B M'_{I,B}} C^{F'_B M'_{F,B}}_{J_B M'_B I_B M'_{I,B}}
         \sum_{M_A  M_{I,A}}  C^{F_A M_{F,A}}_{J_A M_A I_A M_{I,A}}
         \sum_{M_B  M_{I,B}}  C^{F_B M_{F,B}}_{J_B M_B I_B M_{I,B}}
  \\ & \cross
    \langle \beta_A J_A M_A' \beta_B J_B M_B' | \hat{V}^{(2)}_\mathrm{AB}
      | \beta_A J_A M_A \beta_B J_B M_B \rangle
    \langle I_A M'_{I,A} | I_A M_{I,A} \rangle
    \langle I_B M'_{I,B} | I_B M_{I,B} \rangle .
  \end{aligned}
\end{equation}
Because $\hat{V}^{(2)}_\mathrm{AB}$ does not act on nuclear spins, $M'_{I,A} = M_{I,A}$ and $M'_{I,B} = M_{I,B}$. Moreover, Eq.~\eqref{Vmatrix} depends on the pairs $(M_A,M'_A)$ and $(M_B,M'_B)$ through the CG coefficients $C^{J_A M_A'}_{J_A M_A k_A q}$ and $C^{J_B M_B'}_{J_B M_B k_B, -q}$. Applying the relation \cite{VMK}
\begin{equation}
  \sum_{M M' M_I} C^{F' M'_F}_{J M' I M_I} C^{J M'}_{J M k q}
    C^{F M_F}_{J M I M_I} = (-1)^{I+F+k+J} \sqrt{(2J+1)(2F+1)}
    \begin{Bmatrix}
      J  & I & F  \\
      F' & k & J 
    \end{Bmatrix}
    C^{F M'_F}_{F M_F k q}
\end{equation}
to $A$ and $B$ separately, one obtains
\begin{equation}
  \label{VmatrixF}
  \begin{aligned}
    & \langle \beta_A J_A I_A F_A' M_{F,A}' \beta_B J_B I_B F_B' M_{F,B}' 
    | \hat{V}^{(2)}_\mathrm{AB} 
    | \beta_A J_A I_A F_A M_{F,A} \beta_B J_B I_B F_B M_{F,B} \rangle 
  \\ & = -\sum_{l_A l_B l_A' l_B' } \frac{
    (-1)^{l_B + l_B' - J_A + I_A + F_A - J_B + I_B + F_B}} {R^{2+l_A+l_B+l_A'+l_B'}} 
    \sqrt{ \frac{(2 l_A + 2 l_B +1)! (2 l_A' + 2 l_B' +1)!}
                {(2 l_A)! (2 l_B)!(2 l_A')! (2 l_B')!}} 
  \\ & \cross\sum_{k_A k_B k q} (2 k_A +1) (2 k_B +1) 
    C^{k 0}_{l_A+l_B,0,l_A'+l_B',0} C^{k 0}_{k_A q k_B -q} 
  \begin{Bmatrix}
    l_B' & l_B & k_B \\
    l_A' & l_A & k_A \\
    l_A'+l_B' & l_A+l_B  & k
  \end{Bmatrix}
  \\ & \cross\sum_{\beta_A'' J_A'' \beta_B'' J_B''}
    \frac{\langle\beta_A J_A \|\hat{Q}_{l_A }\| \beta_A'' J_A'' \rangle
      \langle\beta_A'' J_A'' \|\hat{Q}_{l_A'}\| \beta_A J_A \rangle
      \langle\beta_B J_B \|\hat{Q}_{l_B}\| \beta_B'' J_B''\rangle
      \langle\beta_B'' J_B'' \|\hat{Q}_{l_B'}\| \beta_B J_B\rangle}
      {E_{\beta_A'' J_A''} - E_{\beta_A J_A}+E_{\beta_B'' J_B''} - E_{\beta_B J_B}} 
  \\ & \cross
  \begin{Bmatrix}
    l_A' & l_A & k_A  \\
    J_A  & J_A & J_A'' 
  \end{Bmatrix} \begin{Bmatrix}
    J_A  & I_A & F_A \\
    F'_A & k_A & J_A' 
  \end{Bmatrix} \begin{Bmatrix}
    l_B' & l_B & k_B \\
    J_B & J_B & J_B'' 
  \end{Bmatrix} \begin{Bmatrix}
    J_B  & I_B & F_B \\
    F'_B & k_B & J'_B 
  \end{Bmatrix} \sqrt{(2 F_A+1) (2 F_B+1)}
    C^{F'_A M'_{F,A}}_{F_A M_{F,A} k_A q} C^{F'_B M'_{F,B}}_{F_B M_{F,B} k_B, -q} \,.
  \end{aligned}
\end{equation}

Next, we apply Eq.~\eqref{VmatrixF} to the van der Waals interaction with $A = \{ \text{Dy}(^5I_8),\, \text{Dy}(^5I_7),\, \text{Er}(^3H_6),\, \text{Er}(^3F_4)\}$ and atom $B = \{\text{AM}(^2S_{1/2}),\, \text{AEM}(^1S_0)\}$. Again, the B atom is only characterized by an isotropic term $k_B = q = 0$ involving the scalar polarizability. This implies $F'_B = F_B$, $M'_{F,A} = M_{F,A}$ and $M'_{F,B} = M_{F,B}$. But, due to the term with $k_A = 2$, different $F_A$ states can be applied, such that $|F'_A - F_A| \le 2$. The van der Waals interaction is thus characterized by possibly off-diagonal $C_6 (F'_A, F_A; F_B, M_{F,A}, M_{F,B})$ coefficients: 
\begin{equation}
  \label{c6f}
  \begin{aligned}
  & C_6 (F'_A, F_A; F_B, M_{F,A}, M_{F,B})
  \\ & = 30 \sqrt{(2 F_A+1) (2 F_B+1)} \sum_{k_A = 0, 2} (2 k_A +1)
    C^{k_A 0}_{2 0 2 0} C^{F_A M_{F,A}}_{F_A M_{F,A} k_A 0} 
    \begin{Bmatrix}
      1 & 1 & 0 \\
      1 & 1 & k_A \\
      2 & 2 & k_A
    \end{Bmatrix} \begin{Bmatrix}
      F_A & F_A & k_A \\
      J_A & J_A & I_A 
    \end{Bmatrix} \begin{Bmatrix}
      F_B & F_B & 0 \\
      J_B & J_B & I_B 
    \end{Bmatrix}
  \\ & \cross \sum_{\beta_A''  J_A'' \beta_B'' J_B''}
    (-1)^{I_A + F_A - J_A'' + I_B + F_B - J_B''}
    |\langle\beta_A J_A \|\hat{Q}_1\| \beta_A'' J_A'' \rangle|^2
    |\langle\beta_B J_B \|\hat{Q}_1\| \beta_B'' J_B'' \rangle|^2
  \\ &\cross \frac{2}{\pi} \int_{0}^{\infty} \mathrm{d}\omega
    \frac{  E_{\beta_A'' J_A''} - E_{\beta_A J_A} }
         { (E_{\beta_A'' J_A''} - E_{\beta_A J_A})^2 + \omega^2 } \cross 
    \frac{  E_{\beta_B'' J_B''} - E_{\beta_B J_B} }
         { (E_{\beta_B'' J_B''} - E_{\beta_B J_B})^2 + \omega^2 }
    \begin{Bmatrix} 
      1   & 1   & k_A \\
      J_A & J_A & J_A'' 
    \end{Bmatrix} \begin{Bmatrix}
      1   & 1   & 0 \\
      J_B & J_B & J_B'' 
    \end{Bmatrix}
  \\ & = C_{6,00} + \frac{(-1)^{I_A + F_A + J_A}}{2} \sqrt{
    \frac{(J_A+1)(2J_A+1)(2J_A+3)(2F_A+1)}{J_A(2J_A-1)} }
    \begin{Bmatrix}
      F_A & F_A' & 2   \\
      J_A & J_A  & I_A 
    \end{Bmatrix} 
    C^{F_A' M_{F,A}}_{F_A M_{F,A} 2 0} C_{6,20}
  \end{aligned}
\end{equation}
\end{widetext}
where $C_{6,00}$ and $C_{6,20}$ are respectively given by Eqs.~\eqref{C60} and \eqref{C62}.
Equation \eqref{c6f} shows that the $C_6$ coefficients form a matrix whose elements are linear combinations of $C_{6,00}$ and $C_{6,20}$, which will be presented in the next section.

\section{Results and discussion}
\label{sec:results}

\subsection{Static polarizabilities}


We start with showing in Table \ref{tab:pola} the scalar and tensor static polarizabilities of the ground and first excited states of Dy and Er. Except for Er$({}^3F_4)$ state, those results have been published in our previous papers \cite{LepersPRA14, LiJPhysB16, BecherPRA18, chalopin2018, patscheider2021}, and have proven to be in satisfactory agreement with literature \cite{rinkleff1994, chu2007, ma2015, dzuba2016}. The small differences with earlier results are due to the fact that here we include into the sum-over-state formulas of Eqs.~\eqref{alphadip}--\eqref{alphatens} the experimental energies where they are known.

\begin{table}[tb!]
	\caption{Scalar and tensor static polarizabilities (in a.u.) of the ground and first excited states of Dy and Er. \label{tab:pola}}
	\begin{ruledtabular}
	\begin{tabular}{cccc}
     Atom & state & $\alpha^\mathrm{scal}$ & $\alpha^\mathrm{tens}$ \\
    \hline
	Dy  &  $^5I_8$  &  163.6  &   1.1  \\
	    &  $^5I_7$  &  163.6  &   1.2  \\ 
	Er  &  $^3H_6$  &  149.3  &  -1.9  \\ 
	    &  $^3F_4$  &  149.1  &   0.1  \\ 
\end{tabular}
\end{ruledtabular}
\end{table}

In addition to static polarizabilities, ac values have also been measured in the context of ultracold gases at the wavelength of 1064 nm. The latter is sufficiently far from absorption resonances, not to depend on a given energetically close transition. In this respect, it resembles polarizabilities at imaginary frequencies which do not possess any peak in wave number due to the $+\omega^2$ term at the denominator. With the present atomic data set, we obtain for ground-state Dy, $\alpha^\mathrm{scal}_{\beta,J} = 192.9$ and $\alpha^\mathrm{tens}_{\beta,J} = 1.5$ a.u., while the experimental values are 184.4 and 1.7 a.u. \cite{RavensbergenPRL18}. For ground-state Er, $\alpha^\mathrm{scal}_{\beta,J} = 173$ and $\alpha^\mathrm{tens}_{\beta,J} = -3.2$ a.u., while the experimental values are 168 and -1.9 a.u.~\cite{BecherPRA18}. For the scalar polarizability, our result is 4.6 and 4.2~\% higher for Dy and Er, respectively. Therefore, we estimate our range of uncertainty on polarizabilities (resp.~$C_6$ coefficients) to be 5~\% of the scalar (resp.~ isotropic) values.

\subsection{$C_6$ coefficients}

In Tables~\ref{tab:tab1}--\ref{tab:tab4}, we present the computed isotropic $C_{6,00}$ and anisotropic $C_{6,20}$ coefficients characterizing the leading term of van der Waals interactions between: Dy($^5I_8$)/Dy($^5I_7$) and alkali-metal AM$(^2S_{1/2})$ atoms (Table~\ref{tab:tab1}), Dy($^5I_8$)/Dy($^5I_7$) and alkaline-earth-metal AEM$(^1S_0)$ atoms (Table~\ref{tab:tab2}), Er($^3H_6$)/Er($^3F_4$) and alkali-metal AM$(^2S_{1/2})$ atoms (Table~\ref{tab:tab3}), and Er($^3H_6$)/Er($^3F_4$) and alkaline-earth-metal AEM$(^1S_0)$ atoms (Table~\ref{tab:tab4}).

For a given lanthanide atom, say Dy$(^5I_8)$, the hierarchy of $C_{6,00}$ coefficients follows the hierarchy of partners' static polarizabilities. Among AMs, $C_{6,00}$ increases with an atomic number except for Fr. For AEMs, $C_{6,00}$ increases with an atomic number from Be to Ba. Moreover, the $C_{6,00}$ coefficients are almost equal for ground-state and first-excited lanthanide atoms, just like the static polarizabilities (see Table~\ref{tab:pola}).

\begin{table}[tb!]
	\caption{$C_6$ coefficients (in a.u.) characterizing the van der Waals interactions of dysprosium atoms in the ground~$^5I_8$ and first excited electronic state~$^5I_7$ with alkali-metal (AM) atoms in the ground electronic state $^2S_{1/2}$. \label{tab:tab1}}
	\begin{ruledtabular}
	\begin{tabular}{lcccc}
	\multirow{2}{*}{AM} & \multicolumn{2}{c}{Dy($^5I_8$) + AM($^2S_{1/2}$)} &\multicolumn{2}{c}{Dy($^5I_7$) + AM($^2S_{1/2}$)} \\
\cline{2-3} \cline{4-5}
	 & $C_{6,00}$ & $C_{6,20}$ & $C_{6,00}$ & $C_{6,20}$\\ 
\hline
	Li  &  1725  &  7.809  &  1725  &  8.104 \\
	Na  &  1850  &  8.033  &  1850  &  8.301 \\ 
	K   &  2857  &  13.13  &  2857  &  13.66 \\ 
	Rb  &  3139  &  14.29  &  3139  &  14.85 \\ 
	Cs  &  3762  &  17.24  &  3763  &  17.94 \\ 
	Fr  &  3372  &  14.47  &  3373  &  14.95 \\ 
\end{tabular}
\end{ruledtabular}
\end{table}

\begin{table}[tb!]
	\caption{$C_6$ coefficients (in a.u.) characterizing the van der Waals interactions of dysprosium atoms in the ground~$^5I_8$ and first excited electronic state~$^5I_7$ with alkaline-earth metal (AEM) atoms in the ground electronic state $^1S_0$. \label{tab:tab2}}
	\begin{ruledtabular}
		\begin{tabular}{lcccc}
			\multirow{2}{*}{AEM} & \multicolumn{2}{c}{Dy($^5I_8$) + AEM($^1S_0$)} &\multicolumn{2}{c}{Dy($^5I_7$) + AEM($^1S_0$)} \\
			\cline{2-3} \cline{4-5}
			& $C_{6,00}$ & $C_{6,20}$ & $C_{6,00}$ & $C_{6,20}$ \\ 
			\hline
			Be  &   671  &  1.997  &   671  &  1.973  \\
			Mg  &  1174  &  3.822  &  1174  &  3.821  \\ 
			Ca  &  2193  &  8.250  &  2193  &  8.393  \\ 
			Sr  &  2651  &  10.21  &  2651  &  10.41  \\ 
			Ba  &  3405  &  13.69  &  3406  &  14.03  \\ 
		\end{tabular}
	\end{ruledtabular}
\end{table}

For all atom pairs, the isotropic coefficients strongly dominate the anisotropic ones, to the same extent as the scalar polarizabilities dominate the tensor ones (see Table \ref{tab:pola}). Note also that the AM- and AEM-Er$({}^3H_6)$ $C_{6,20}$ coefficients are all negative, exactly like $\alpha^\mathrm{tens}_{^3H_6}$. Finally, the $C_{6,20}$ coefficients involving Er$({}^3F_4)$ are at least one order of magnitude smaller than those involving Er$({}^3H_6)$, in consistency with the ratio $-1/19$ of their tensor polarizabilities. This surprising result is likely to come from the peculiar nature of Er's first excited state, namely $68~\% \, ({}^3F_4) + 25~\% \, ({}^1G_4)$; it could be confirmed by a measurement of its static or 1064-nm tensor polarizability.

\begin{table}[tb!]
	\caption{$C_6$ coefficients (in a.u.) characterizing the van der Waals interactions of erbium atoms in the ground~$^3H_6$ and first excited electronic state~$^3F_4$ with alkali-metal (AM) atoms in the ground electronic state $^2S_{1/2}$. \label{tab:tab3}}
	\begin{ruledtabular}
		\begin{tabular}{lcccc}
			\multirow{2}{*}{AM} & \multicolumn{2}{c}{Er($^3H_6$) + AM($^2S_{1/2}$)} &\multicolumn{2}{c}{Er($^3F_4$) + AM($^2S_{1/2}$)} \\
			\cline{2-3} \cline{4-5}
			& $C_{6,00}$ & $C_{6,20}$ & $C_{6,00}$ & $C_{6,20}$\\ 
			\hline
			Li & 1609 & -9.210 & 1607 & 0.6894\\
			Na & 1729 & -8.956 & 1727 & 0.6578 \\ 
			K & 2664 & -15.89  & 2661 & 1.169 \\ 
			Rb & 2929 & -17.13 & 2925 &  1.242 \\ 
			Cs & 3509 & -20.95 & 3506 & 1.501 \\ 
			Fr & 3156 & -16.14 & 3152 &  1.105 \\ 
		\end{tabular}
	\end{ruledtabular}
\end{table}

\begin{table}[tb!]
	\caption{$C_6$ coefficients (in a.u.) characterizing the van der Waals interactions of erbium atoms in the ground~$^3H_6$ and first excited electronic state~$^3F_4$ with alkaline-earth metal (AEM) atoms in the ground electronic state $^1S_0$. \label{tab:tab4}}
	\begin{ruledtabular}
		\begin{tabular}{lcccc}
			\multirow{2}{*}{AEM} & \multicolumn{2}{c}{Er($^3H_6$) + AEM($^1S_0$)} &\multicolumn{2}{c}{Er($^3F_4$) + AEM($^1S_0$)} \\
			\cline{2-3} \cline{4-5}
			& $C_{6,00}$ & $C_{6,20}$ & $C_{6,00}$ & $C_{6,20}$\\ 
			\hline
			Be  &   636  &  -0.8894 &  636  &  0.0029  \\
			Mg  &  1110  &  -2.367  & 1109  &  0.0950  \\ 
			Ca  &  2063  &  -7.279  & 2061  &  0.4608  \\ 
			Sr  &  2491  &  -9.435  & 2489  &  0.6117  \\ 
			Ba  &  3195  & -13.68   & 3192  &  0.9187  \\ 
		\end{tabular}
	\end{ruledtabular}
\end{table}

\section{Summary and conclusions}
\label{sec:summary}

In the present work, we have provided analytical expressions for the isotropic and anisotropic $C_6$ van der Waals coefficients for the interaction between a non-$S$-state atom and an $S$-state atom, including the atomic hyperfine quantum numbers. We have applied the derived formulas to compute the $C_6$ coefficients for the Dy($^5I_8$)/Dy($^5I_7$)/Er($^3H_6$)/Er($^4F_3$) + AM($^2 S_{1/2}$)/AEM($^1 S_0$) systems, where AM = Li, Na, K, Rb, Cs, Fr and AEM = Be, Mg, Ca, Sr, Ba. For all atom pairs, the isotropic $C_6$ coefficients are two or three orders of magnitude larger than the anisotropic ones, following the hierarchy between scalar and tensor static polarizabilities. Those results are similar to those obtained for Dy-Dy \cite{LiJPhysB16} and Er-Er pairs \cite{LepersPRA14}.

With the rapid developments in the field of dipolar quantum gases and ongoing experiments involving highly magnetic lanthanide atoms, the present results will be beneficial for studies of collisional properties of heteronuclear quantum mixtures containing dysprosium or erbium atoms and alkali-metal or alkaline-earth metal atoms. In particular, the anisotropic $C_{6,20}$ coefficients are expected to be the main source of coupling between scattering channels, and so of the emergence of Feshbach resonances in those systems. The derived formulas can also be employed to calculate long-range coefficients for other similar neutral or ionic atomic combinations.

\begin{acknowledgments}

K.~Z.-K. acknowledges the financial support from the National Science Center, Poland (grant no.~2019/35/N/ST4/04504). The work reported here was initiated during K. Z.-K.'s visit at the Universit\'{e} de Bourgogne Franche-Comt\'{e} funded by the Excellence Initiative -- Research University Program. M.~T. acknowledge the financial support from the Foundation for Polish Science within the First Team program co-financed by the European Union under the European Regional Development Fund and Poland's high-performance computing infrastructure PLGrid (HPC Center: ACK Cyfronet AGH) for providing computer facilities and support (computational grant no.~PLG/2023/016115).

\end{acknowledgments}


%

\end{document}